\documentclass[prb,notitlepage,reprint]{revtex4-1}

\usepackage{amsmath}
\usepackage{amsfonts}
\usepackage{graphicx}
\usepackage{color}
\usepackage{float}
\usepackage{leftindex}
\usepackage{gensymb}
\usepackage[dvipsnames]{xcolor}

\begin{document}

\title{Non-Equilibrium Sock Dynamics: Spontaneous Symmetry Breaking in the Agitated Wash}

\date{April 1st 2026}

\author{Ahmad Darwish}
\author{Matteo Murdaca}
\author{Jami J. Kinnunen}
\email{jami.kinnunen@aalto.fi}
\affiliation{Department of Applied Physics, Aalto University School of Science, FI-00076 Aalto, Finland}

\begin{abstract}
It is a universal empirical observation that socks become unpaired in the laundry. We propose a quasiparticle theory of sock dynamics in which individual socks are modelled as bosonic excitations of the agitated laundry condensate. The sock dispersion relation is material-dependent: nondispersive materials retain their shape, while dispersive materials give rise to the well-documented phenomenon of sock shrinkage. In the convex regions of the dispersive spectrum, socks undergo Beliaev decay and spontaneously split into two lower-momentum socks, while in the concave regions the dominant process is Landau--Khalatnikov scattering, which degrades socks into lint and loose threads. In addition, the rotating drum creates sock--antisock pairs from the laundry vacuum via the dynamical Casimir effect. The coexistence of these creation and destruction channels gives rise to a fundamental ambiguity: an unpaired sock at the end of a wash cycle is equally consistent with the destruction of its partner or the spontaneous creation of an entirely new sock.
\end{abstract}

\maketitle

\section{Introduction}

\begin{figure}[ht]
    \centering
    \includegraphics[width=0.95\columnwidth]{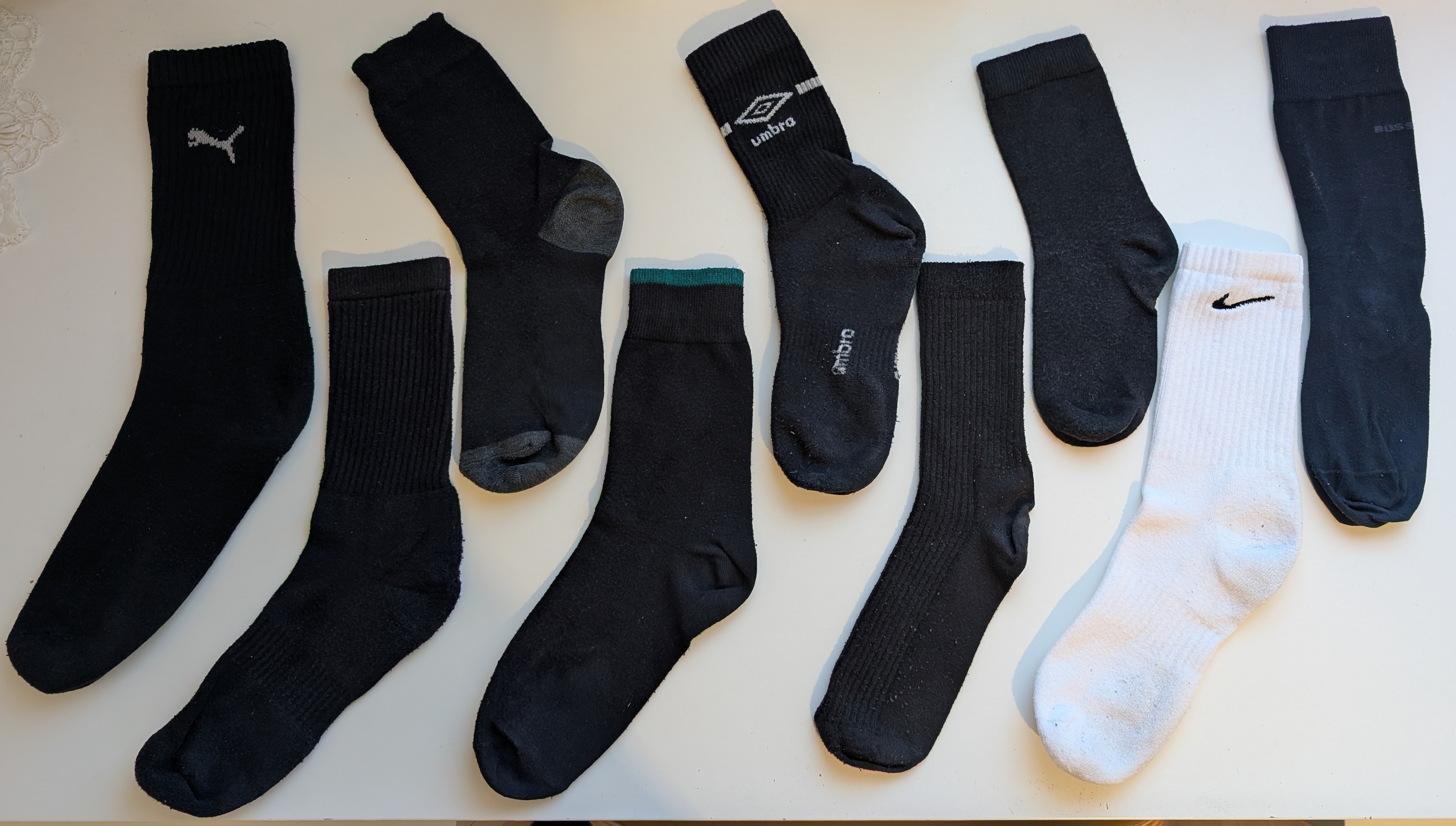}
    \caption{Nine unpaired sock quasiparticles recovered from a domestic laundry system over a period of six months. The sample exhibits a broad distribution of sizes, colors, and brands, consistent with multiple decay and creation channels contributing to the unpaired population. The predominance of dark socks is noted but not yet understood within the present theoretical framework (see Sec.~VI for discussion). The variety of brands (Puma, Nike, Umbro, Boss, and unbranded) suggests that the unpairing process is universal and does not depend on the manufacturer.}
    \label{fig:unpaired-socks}
\end{figure}

The disappearance of socks in the laundry is one of the most widely documented yet poorly understood phenomena in domestic physics. Despite decades of empirical observation (for latest experiment see Fig.~\ref{fig:unpaired-socks}), no consensus exists on the underlying mechanism. Proposed explanations range from mechanical entrapment in the drum seals to spontaneous teleportation~\cite{TODO-teleportation}. None of these theories have provided a quantitative prediction for the rate of sock loss.

In this paper, we approach the problem from the perspective of many-body quantum field theory. The key insight is that the laundry, consisting of a large number of interacting garments undergoing collective agitated motion, constitutes a many-body system whose low-energy excitations can be described by quasiparticles. Individual socks, being distinguishable from the bulk of the laundry (shirts, towels, etc.), are naturally identified as quasiparticle excitations of the laundry condensate.

A crucial property of any quasiparticle is its dispersion relation $\varepsilon(\mathbf{p})$, which determines the energy as a function of momentum. The shape of the dispersion --- in particular whether it is concave or convex --- determines which decay processes are kinematically allowed. For a convex dispersion, the Beliaev process~\cite{beliaev}, in which a single quasiparticle splits into two quasiparticles of lower energy, is allowed by energy and momentum conservation. For a concave dispersion, Beliaev decay is forbidden, and the dominant dissipation mechanism is the Landau--Khalatnikov process~\cite{khalatnikov}, in which a quasiparticle scatters off thermal excitations and is degraded into lower-lying modes.

We derive the sock quasiparticle dispersion relation and show that it features a characteristic minimum at a wavevector corresponding to the inverse sock length --- the \emph{sockton minimum}, in analogy with the roton minimum in superfluid helium~\cite{landau-roton}. The velocity gradient inside the rotating drum naturally provides the momentum hierarchy required for Beliaev decay: a high-momentum sock near the outer drum wall can split into two \mbox{lower-momentum} socks in the interior.

Crucially, we note that the fundamental observable is not the \emph{disappearance} of a sock but rather the appearance of an \emph{unpaired} sock. This distinction is essential, as both the Beliaev process and a dynamical Casimir mechanism~\cite{casimir-dynamical} driven by the rotating drum \emph{create} socks, while only the Landau--Khalatnikov process destroys them. The resulting creation--annihilation ambiguity means that an unpaired sock may signal either the loss of its partner or the creation of an entirely new sock. We show that the rates of all three processes depend on the drum frequency $\Omega$, establishing a direct connection between the spin cycle speed and the rate of sock unpairing.

\section{The laundry as a many-body system}






The full many-body Hamiltonian in the rotating reference frame is
$$
\hat H = \sum_{k\sigma} \varepsilon_{k\sigma} a_{k\sigma}^\dagger a_{k\sigma} + g \sum_{kpq,\sigma,\sigma'} a_{k\sigma}^\dagger {a_{q-k,\sigma'}^\dagger a_{q-p,\sigma'} a_{p}} - \Omega \cdot \hat {\bf L},
$$
where $\Omega$ is the angular frequency of the laundry machine drum. Interactions are described by contact interaction potential with well-known scattering properties~\cite{Griffiths}. 

The creation and annihilation operators $a_{k\sigma}^\dagger,a_{k\sigma}$ operate on the underlying many-body laundry field, but the actual quasiparticle sock excitations are obtained as bosonic excitations of the field. At the mean-field level, the quasiparticle Hamiltonian becomes
$$
\hat H' = \sum_{k\sigma} \epsilon_{k\sigma} \gamma_{k\sigma}^\dagger \gamma_{k\sigma},
$$
where $\epsilon_{k\sigma}$ is the sock quasiparticle dispersion relation. A typical dispersion relation is depicted in Fig.~\ref{fig:dispersion} with a striking resemblance to the quasiparticle excitations in superfluid $^4$He.

Notice that the quasiparticle Hamiltonian describes quasiparticles as noninteracting excitations. This is due to the mean-field approximation which neglects residual interactions between quasiparticles that arise from beyond-mean-field effects. It is these effects that ultimately provide the decay mechanisms for the socks observed in domestic rotation experiments.

\section{Sock quasiparticle dispersion relation}

\begin{figure*}[ht]
    \centering
    \includegraphics[width=0.9\textwidth]{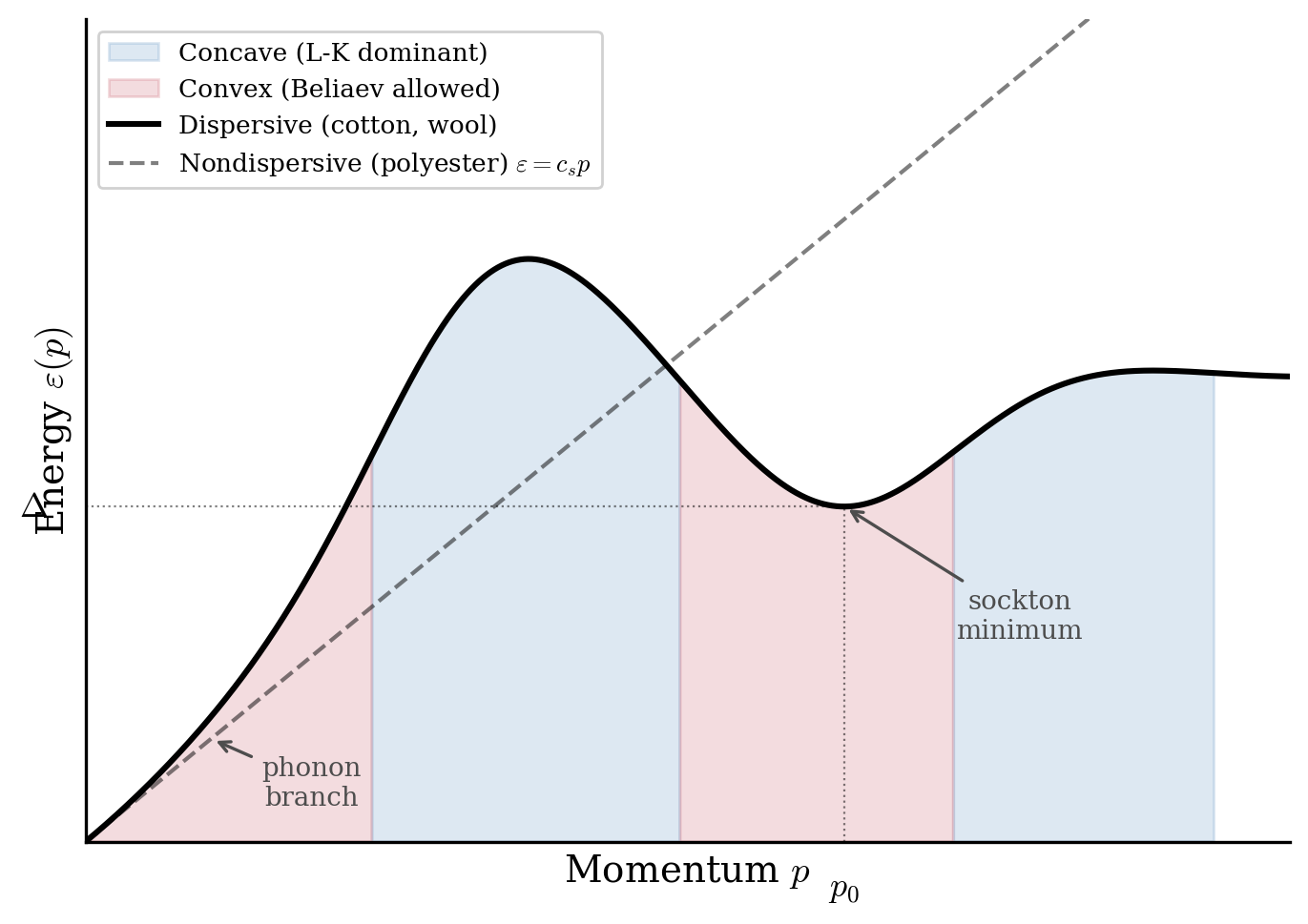}
    \caption{The sock quasiparticle dispersion relation $\varepsilon(p)$ for dispersive (solid) and nondispersive (dashed) sock materials. The dispersive spectrum exhibits a linear phonon branch at low momenta, a convex region where Beliaev decay is kinematically allowed (red shading), and a sockton minimum at momentum $p_0$ with energy gap $\Delta$, surrounded by a concave region where the Landau--Khalatnikov process dominates (blue shading). The nondispersive (linear) spectrum $\varepsilon = c_s p$ characteristic of synthetic materials is shown for comparison.}
    \label{fig:dispersion}
\end{figure*}

The sock quasiparticle dispersion relation is obtained by considering the spectrum of collective excitations in the agitated laundry medium and a typical relation is shown in Fig.~\ref{fig:dispersion}.

The spectrum is gapless, with the long wavelength excitations approaching zero energy. The gaplessness is a direct consequence of the Goldstone theorem~\cite{goldstone}: when a continuous symmetry is spontaneously broken, the spectrum of excitations necessarily contains a massless (gapless) mode --- the Nambu--Goldstone boson. Notice the lack of Higgs mechanism~\cite{higgs,anderson-higgs}: sock-sock interactions are contact interaction and thus have not gauge field structure. The laundry condensate is therefore a \emph{neutral superfluid}, analogous to superfluid $^4$He or an ultracold atomic Bose--Einstein condensate, in which only global symmetries are broken. 

 However, depending on material parameters, the linear part of the spectrum may hold only in the very long wavelength limit, with the relation becoming dispersive. At shorter wavelengths, an energy minimum may appear. This is called the sockton minimum, as it is an analogue of the roton minimum in the superfluid $^4$He spectrum. 

\subsection{Material dependence of the dispersion}

The dispersion relation of the sock quasiparticle depends critically on the material composition of the sock. We can classify sock materials into two broad categories based on their spectral properties.

For certain synthetic materials (e.g.\ polyester, nylon), the low-energy excitation spectrum is approximately \emph{nondispersive}, i.e.\ linear in momentum:
\begin{equation}
    \varepsilon(\mathbf{p}) \approx c_s |\mathbf{p}|,
\end{equation}
where $c_s$ is the speed of sock in the medium. For a linear dispersion, the group velocity $v_g = \partial \varepsilon / \partial p = c_s$ is independent of momentum, and consequently wave packets propagate without spreading or deformation. The sock quasiparticle retains its spatial profile indefinitely. This is consistent with the empirical observation that synthetic socks maintain their shape and size over many wash cycles.

In contrast, natural fibers (e.g. cotton, wool) give rise to a strongly \emph{dispersive} spectrum, where the group velocity $v_g = \partial \varepsilon / \partial p$ varies significantly with momentum. A dispersive spectrum causes wave packets to deform over time. The macroscopic consequence is immediately recognizable: \emph{sock shrinkage}. The well-documented phenomenon of cotton and wool socks shrinking in the wash at high temperatures and particularly in temperature quenches constitutes direct experimental evidence for a dispersive sock quasiparticle spectrum. The susceptibility to spatial deformation is quantified by spatial curvature $\kappa$, defined as $\partial^2\varepsilon/\partial p^2$. Socks made of natural fibers possess a non-zero $\kappa$, which leads to a finite effective sock mass $m^* = \hbar^2/\kappa$. In the many-body system of a sock in the laundry, the effective mass of the sock describes its shrinkage response. Socks with a heavy effective mass resist spatial deformation and are less prone to shrinkage. On the other hand, socks with a lighter effective mass experience strong dispersion, and accordingly, rapidly shrink. 




\section{Decay processes and sock lifetime}

\begin{figure*}[ht]
    \centering
    \includegraphics[width=\textwidth]{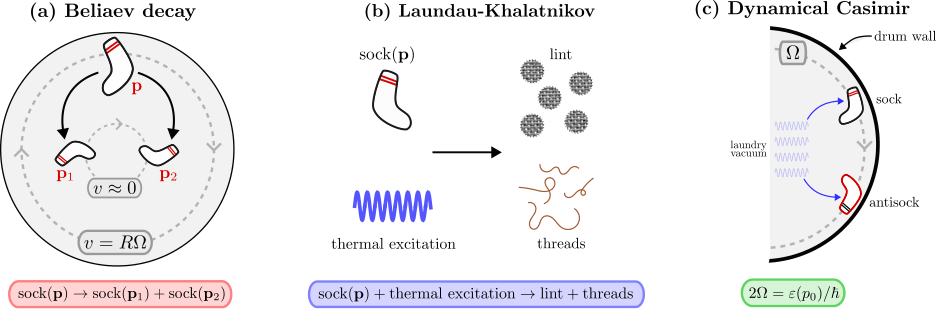}
    \caption{Schematic illustration of the three mechanisms governing the sock population. (a)~Beliaev decay: a high-momentum sock near the outer drum wall (velocity $v = R\Omega$) splits into two lower-momentum socks in the drum interior. (b)~Landau--Khalatnikov process: a sock scatters off a thermal excitation and is degraded into lint and loose threads. (c)~Dynamical Casimir effect: the accelerating drum wall creates a sock--antisock pair from the laundry vacuum when the resonance condition $2\Omega = \varepsilon(p_0)/\hbar$ is satisfied.}
    \label{fig:three-processes}
\end{figure*}

The dispersion relation is closely related to quasiparticle decay mechanisms~\cite{castin}. In particular, Beliaev decay is possible in the region where the dispersion relation is convex, whereas for a concave relation the dominant decay mechanism is the Landau-Khalatnikov process.

Due to the interplay between the dispersivity of the spectrum and the decay mechanisms it is clear that materials with dispersive spectrum are also the ones susceptible to Beliaev decay. So shrinking socks are also the ones most likely to vanish, which is an experimentally testable property.





\subsection{Beliaev decay}

In the convex region of the dispersion, the dominant process is Beliaev decay~\cite{beliaev}, in which a sock quasiparticle at momentum $\mathbf{p}$ spontaneously splits into two sock quasiparticles of lower energy:
\begin{equation}
    \mathrm{sock}(\mathbf{p}) \rightarrow \mathrm{sock}(\mathbf{p}_1) + \mathrm{sock}(\mathbf{p}_2),
\end{equation}
subject to the conservation laws
\begin{align}
    \mathbf{p} &= \mathbf{p}_1 + \mathbf{p}_2, \\
    \varepsilon(\mathbf{p}) &= \varepsilon(\mathbf{p}_1) + \varepsilon(\mathbf{p}_2).
\end{align}
The physical setting for this process is provided by the velocity gradient inside the rotating drum. In the frame of the drum, garments near the outer wall move at the drum's peripheral velocity $v = R\Omega$, while garments in the interior of the laundry mass move at progressively lower velocities. A sock quasiparticle residing in the high-momentum outer layer can therefore decay into two daughter socks that populate the lower-momentum inner layers of the rotating laundry, where the conservation laws are naturally satisfied.

Crucially, the Beliaev process \emph{increases} the total number of socks: one sock becomes two. Unless the parent sock belonged to an already-unpaired population, the two daughter socks will in general not form a matching pair with any other sock in the system. Beliaev decay thus provides a natural mechanism for the generation of unpaired socks.


\subsection{Landau--Khalatnikov process}

In the concave region of the dispersion, Beliaev decay is kinematically forbidden. Instead, the dominant mechanism is the Landau--Khalatnikov process~\cite{khalatnikov}, in which a sock quasiparticle scatters off thermal excitations of the laundry medium and is degraded into non-sock excitations:
\begin{equation}
    \mathrm{sock}(\mathbf{p}) + \mathrm{thermal}(\mathbf{p}') \rightarrow \mathrm{lint}(\mathbf{p}_1) + \mathrm{threads}(\mathbf{p}_2).
\end{equation}
Unlike Beliaev decay, this process genuinely \emph{destroys} the sock, converting it into lint and loose threads. The rate of the Landau--Khalatnikov process is proportional to the thermal occupation of excitations in the laundry medium, and is therefore strongly enhanced at higher wash temperatures. The remnants of the destroyed sock are routinely recovered in the lint trap, providing direct experimental confirmation of this dissipation channel.


\subsection{Sock creation by the dynamical Casimir effect}

In addition to Beliaev decay, the rotating drum provides a second, independent mechanism for sock creation. The drum wall constitutes a periodically accelerating boundary condition for the sock field. In quantum field theory, it is well established that accelerating boundaries can create particle pairs from the vacuum --- the dynamical Casimir effect~\cite{casimir-dynamical, wilson-casimir}. By direct analogy, the rotating drum can create sock--antisock pairs from the laundry vacuum.

%

The creation rate is governed by the Bogoliubov coefficient relating the sock vacuum states before and after the spin cycle. For a drum rotating at angular frequency $\Omega$, resonant pair creation occurs when
\begin{equation}
    2\Omega = \frac{\varepsilon(p_0)}{\hbar},
    \label{eq:resonance}
\end{equation}
where $p_0$ is the momentum at the sockton minimum. This resonance condition establishes a direct, testable relationship between the spin cycle speed and the rate of sock creation, and together with the Beliaev and Landau--Khalatnikov rates determines the net sock balance of each wash cycle.


\section{The creation--annihilation ambiguity}

The results of the preceding sections reveal a fundamental ambiguity in the interpretation of the sock problem. Both the Beliaev process (Sec.~IV\,A) and the dynamical Casimir mechanism (Sec.~IV\,C) \emph{increase} the total number of socks, while the Landau--Khalatnikov process (Sec.~IV\,B) \emph{decreases} it. The experimentally accessible observable --- the appearance of unpaired socks --- is consistent with either net creation or net destruction, and the two cannot be distinguished without a careful sock census.

Consider a wash cycle that begins with an even number $2n$ of socks and ends with an odd number $2n \pm 1$. The parity change establishes that something has happened, but the sign of the change is ambiguous: the observation is equally consistent with the destruction of one sock (Landau--Khalatnikov, total $2n - 1$) and the creation of one sock (Beliaev decay or dynamical Casimir effect, total $2n + 1$). We note that such census measurements are rarely performed in domestic settings, leaving the creation--annihilation ambiguity unresolved in most practical situations.


%

%
%

\section{Conclusions}

We have developed a quasiparticle theory of sock dynamics in the laundry, deriving the sock dispersion relation from the many-body Hamiltonian of the agitated laundry condensate. The theory identifies three distinct mechanisms that govern the sock population during a wash cycle: Beliaev decay, in which a high-momentum sock near the drum wall splits into two lower-momentum socks in the drum interior; the Landau--Khalatnikov process, in which a sock is scattered by thermal excitations and degraded into lint and loose threads; and the dynamical Casimir effect, in which the accelerating drum wall creates sock--antisock pairs from the laundry vacuum. Together, these mechanisms give rise to a creation--annihilation ambiguity: the observation of an unpaired sock after a wash cycle is equally consistent with the destruction of its partner and the creation of an entirely new sock.

The theory makes several experimentally testable predictions. First, the material dependence of the dispersion relation implies that socks made of nondispersive (synthetic) materials should be immune to both shrinkage and decay, as their linear spectrum supports neither wave packet deformation nor Beliaev splitting. This is consistent with the empirical longevity of polyester socks compared to their cotton and wool counterparts. Second, the Landau--Khalatnikov rate is proportional to the thermal occupation of the laundry medium, predicting that higher wash temperatures accelerate sock destruction --- a well-known fact among careful launderers. Third, the resonance condition $2\Omega = \varepsilon(p_0)/\hbar$ in Eq.~\eqref{eq:resonance} predicts a critical spin cycle frequency above which sock pair creation is resonantly enhanced, suggesting that lower spin speeds should reduce the rate of sock unpairing.

On the basis of these findings, we offer the following practical recommendations for extending sock lifetime: use synthetic socks with a nondispersive spectrum, wash at low temperatures to suppress the Landau--Khalatnikov channel, and select a low spin speed to remain below the Casimir resonance. We note with some satisfaction that these prescriptions coincide with standard textile care guidelines, lending indirect support to the theory.

Several open questions remain. The nature of the antisock --- whether it corresponds to the parity-inverted (inside-out) state of an ordinary sock, and whether sock--antisock annihilation can be observed under controlled conditions --- deserves further investigation. The role of sock color is also not yet understood within the present framework; the predominance of dark socks in the unpaired population (Fig.~\ref{fig:unpaired-socks}) may hint at a color-dependent coupling constant, but we leave this to future work. Furthermore, it would be of great interest to determine whether the sock dispersion can be engineered to be everywhere convex, yielding topologically protected socks that are immune to Landau--Khalatnikov degradation. Such socks would still undergo Beliaev splitting, however, so they would proliferate rather than vanish --- a different but perhaps equally troublesome domestic outcome. Finally, it is possible to investigate whether the formation of mysterious toe holes in socks can similarly be explained by the dissipative mechanisms discussed in this paper. Sock holes suggests the existence of sock excitons. The formation process and physical realization of such sock excitons continues to be a mysterious aspect of the sock quasiparticle model, requiring further investigation.


\end{document}